\documentstyle[aps,prd,tighten,epsf]{revtex}
\begin{document}
\draft
\newcommand{\be}{\begin{equation}}
\newcommand{\ee}{\end{equation}}
\newcommand{\ben}{\begin{eqnarray}}
\newcommand{\een}{\end{eqnarray}}

\newcommand{\la}{{\lambda}}
\newcommand{\Om}{{\Omega}}
\newcommand{\ta}{{\tilde a}}
\newcommand{\bg}{{\bar g}}
\newcommand{\bh}{{\bar h}}
\newcommand{\si}{{\sigma}}
\newcommand{\th}{{\theta}}
\newcommand{\C}{{\cal C}}
\newcommand{\D}{{\cal D}}
\newcommand{\cA}{{\cal A}}
\newcommand{\cT}{{\cal T}}
\newcommand{\cO}{{\cal O}}
\newcommand{\eeo}{\cO ({1 \over E})}
\newcommand{\G}{{\cal G}}
\newcommand{\cL}{{\cal L}}
\newcommand{\cH}{{\cal H}}
\newcommand{\cE}{{\cal E}}
\newcommand{\M}{{\cal M}}

\newcommand{\p}{\partial}
\newcommand{\na}{\nabla}
\newcommand{\ssum}{\sum\limits_{i = 1}^3}
\newcommand{\dssum}{\sum\limits_{i = 1}^2}
\newcommand{\tal}{{\tilde \alpha}}

\newcommand{\tp}{{\tilde \phi}}
\newcommand{\tPhi}{\tilde \Phi}
\newcommand{\tpsi}{\tilde \psi}
\newcommand{\tim}{{\tilde \mu}}
\newcommand{\tr}{{\tilde \rho}}
\newcommand{\tir}{{\tilde r}}
\newcommand{\rp}{r_{+}}
\newcommand{\hr}{{\hat r}}
\newcommand{\rv}{{r_{v}}}
\newcommand{\dr}{{d \over d \hr}}
\newcommand{\dR}{{d \over d R}}

\newcommand{\hhf}{{\hat \phi}}
\newcommand{\hhM}{{\hat M}}
\newcommand{\hhQ}{{\hat Q}}
\newcommand{\hht}{{\hat t}}
\newcommand{\hhr}{{\hat r}}
\newcommand{\hhS}{{\hat \Sigma}}
\newcommand{\hhD}{{\hat \Delta}}
\newcommand{\hhm}{{\hat \mu}}
\newcommand{\hro}{{\hat \rho}}
\newcommand{\hhz}{{\hat z}}

\newcommand{\tD}{{\tilde D}}
\newcommand{\tB}{{\tilde B}}
\newcommand{\tV}{{\tilde V}}
\newcommand{\hT}{\hat T}
\newcommand{\tF}{\tilde F}
\newcommand{\tT}{\tilde T}
\newcommand{\hC}{\hat C}
\newcommand{\ep}{\epsilon}
\newcommand{\bep}{\bar \epsilon}
\newcommand{\ppp}{\varphi}
\newcommand{\Ga}{\Gamma}
\newcommand{\ga}{\gamma}
\newcommand{\hth}{\hat \theta}
\title{Black Rings and the Physical Process Version of the First Law of Thermodynamics}

\author{Marek Rogatko}
\address{Institute of Physics \protect \\
Maria Curie-Sklodowska University \protect \\
20-031 Lublin, pl.Marii Curie-Sklodowskiej 1, Poland \protect \\
rogat@tytan.umcs.lublin.pl \protect \\
rogat@kft.umcs.lublin.pl}
\date{\today}
\maketitle
\smallskip
\bigskip
\begin{abstract}
We consider the problem of {\it physical process} version of the first law black ring thermodynamics
in n-dimensional Einstein gravity with additional (p+1)-form field strength and dilaton fields. 
The first order variations of mass, angular momentum and local charge for black ring were derived.
By means of them we prove {\it physical process} version of the first law of thermodynamic for stationary
black rings.
\end{abstract}
\baselineskip=18pt
\par
\section{Introduction}
During the past decade there has been growing interest in higher dimensional black holes motivated by 
various attempts of building unified theories. It has been recently shown that for static
$n$-dimensional black holes the uniqueness theorem is valid \cite{uniq}, while for the stationary axisymmetric
ones even in five-dimensional spacetime there is a counterexample, black ring \cite{emp02}. The black ring solution
with the topology horizon of $S^2 \times S^1$ has the same mass and angular momentum as spherical black hole.
Nevertheless, when one assumes the spherical topology of the horizon $S^3$ one can prove the uniqueness
of vacuum five-dimensional stationary axisymmetric solution \cite{mor04} as well as stationary axisymmetric
self-gravitating $\sigma$-models \cite{rog04a}.\\
The black ring solutions can be extended to possess the electric as well as magnetic dipole charge \cite{elv03,elv03b}.
The static black ring solution in five-dimensional Einstein-Maxwell-dilaton gravity was presented in Ref.\cite{kun05}.
In Ref.\cite{elv05} the solution characterized by three conserved charges, three dipole charges, two 
equal angular momentum and a parameter that measured the deviation from the supersymmetry configuration was presented.
It also turned out that supersymmetric black rings exist \cite{sup}.
\par
The first connection between black holes and thermodynamics was presented in the seminal paper of
Bardeen, Carter and Hawking \cite{bar73}. They considered linear
perturbations of a stationary electrovac black hole to another stationary black hole.
Sudarsky and Wald \cite{sud92}  
derived the first law of black hole thermodynamics
valid for arbitrary asymptotically flat perturbations of a stationary 
black hole. There have been several derivations of the first law of black hole thermodynamics valid for
an arbitrary diffeomorphism invariant Lagrangians 
with metric and matter fields possessing stationary and axisymmetric
black hole solutions \cite{wal}.
The higher curvature terms and higher derivative terms in the metric
were considered in
Ref.\cite{jac}, as well as the Lagrangian being arbitrary function of metric,
Ricci tensor and a scalar field \cite{kog98} were taken into account. The case of a charged and rotating black hole
where fields were not smooth through the event horizon was treated in Ref.\cite{gao03}.
On the other hand, the {\it physical process} version of the first law of black hole thermodynamics
obtained by changing a stationary black hole by some infinitesimal physical process, e.g.,
when matter was thrown into black hole was considered. Assuming that the black hole eventually settle down to a 
stationary state and calculating the changes of black hole's parameters one can find this law.
If the resulting relation fails comparing to the known version of 
the first law of black hole thermodynamics it provides inconsistency with the assumption
that the black hole settles down to a final stationary state. This fact will give
a strong evidence against cosmic censorship. The {\it physical process } version of the first 
law of black hole thermodynamics in Einstein theory was proved in Ref.\cite{wal94}. Then, it was
generalized for Einstein-Maxwell (EM) black holes in Ref.\cite{gao01} and for 
Einstein-Maxwell axion-dilaton (EMAD) gravity black holes in
\cite{rog02}.\\
The first law of black hole thermodynamics was also intensively studied in the realm of $n$-dimensional
black holes. The {\it equilibrium state} version was elaborated in Ref.\cite{equi1} 
under the assumption of spherical topology of black holes. Some of the works assume that four-dimensional black 
hole uniqueness theorem extends to higher dimensional case \cite{equi2}.
The {\it physical process} of the first law of black hole thermodynamics in $n$-dimension was 
treated in Ref.\cite{rog05}.
\par
One hopes that the analysis of the physical processes in the spacetime of black rings will deepen our understanding
of this objects. Recently, several works were devoted to this problem. Namely,
the process of Penrose extraction was analyzed \cite{noz05} and
ultrarelativistic boost was taken into account \cite{ort05}. The scalar perturbations in the 
background of both nonsupersymmetric and supersymmetric black rings was considered in Ref.\cite{car05}.
The general form of the first law of mechanics for black rings, taking into account dipole charges
was established in Ref.\cite{cop05}.
\par
In our paper we shall investigate the problem of {\it physical process} version of the first law of thermodynamics
in the higher dimensional gravity containing $(p + 1)$-form field strength and dilaton fields. This theory
constitutes the simplest generalization of five-dimensional one, which in turn contains stationary 
black ring solution with dipole charge \cite{elv05}.

\section{Physical process version of the first law of black hole mechanics}
We begin with the Lagrangian of higher dimensional generalization of
the five-dimensional theory with three form field strength and dilaton fields which contains stationary 
black ring solution.
It is subject to the relation as follows:
\be
{\bf L } = {\bf \ep} \bigg(
{}^{(n)}R - {1 \over 2} \na_{\mu}\phi \na^{\mu} \phi - {1\over 2 (p + 1)!} e^{-{\alpha} \phi}
H_{\mu_{1} \dots \mu_{p+1}} H^{\mu_{1} \dots \mu_{p+1}}
\bigg),
\label{lag}
\ee
where by $ {\bf \ep}$ we denote the volume element,
$\phi$ is the dilaton field while
$H_{\mu_{1} \dots \mu_{p+1}} = (p + 1)! \na_{[ \mu_{1}} B_{{\mu_{2} \dots \mu_{p+1}]}} $ is $(p + 1)$-form field strength.
The equations of motion for the underlying theory imply
\ben \label{m1}
G_{\mu \nu} - T_{\mu \nu}(B, \phi) &=& 0, \\
\na_{j_{1}}\bigg( e^{-{\alpha} \phi} H^{j_{1} \dots j_{p+1}} \bigg) &=& 0,\\ \label{m2}
\na_{\mu} \na^{\mu} \phi + {\alpha \over 2 (p + 1)!} e^{-{\alpha} \phi}
H_{\mu_{1} \dots \mu_{p+1}} H^{\mu_{1} \dots \mu_{p+1}} &=& 0,
\label{m3}
\een
while the energy momentum tensor for $(p + 1)$-form field strength and dilatons has the form as
\ben
T_{\mu \nu}(B, \phi) &=&
{1 \over 2} \na_{\mu} \phi \na_{\nu} \phi - {1 \over 4} g_{\mu \nu} \na_{\alpha} \phi
\na^{\alpha} \phi \\ \nonumber
 &+& {1\over 2 (p + 1)!} e^{-{\alpha} \phi}
\bigg[ (p + 1) H_{\mu \nu_{2} \dots \nu_{p+1}} H_{\nu}{}{}^{\nu_{2} \dots \nu_{p+1}}
- {1 \over 2} g_{\mu \nu} H_{\mu_{1} \dots \mu_{p+1}} H^{\mu_{1} \dots \mu_{p+1}}
\bigg].
\een
In order to find the {\it physical version} of the first law of black rings thermodynamics 
we shall try to find the explicit expressions
for the variation of mass and angular momentum.
To begin with we perform variation of the Lagrangian (\ref{lag}). On evaluating
the variations of the adequate fields, we find that one finally left with
\ben \label{dl}
\delta {\bf L} &=& {\bf \epsilon} \bigg(
G_{\mu \nu} - T_{\mu \nu}(B, \phi) \bigg)~ \delta g^{\mu \nu}
- {\bf \epsilon} \na_{j_{1}} \bigg( e^{-{\alpha} \phi}H^{j_{1} \dots j_{p+1}} \bigg) 
\delta B_{j_{2} \dots j_{p+1}}  \\ \nonumber
&+& {\bf \epsilon} \bigg( \na_{\mu}\na^{\mu} \phi + {\alpha \over 2 (p + 1)!} e^{-{\alpha} \phi}
H_{\mu_{1} \dots \mu_{p+1}} H^{\mu_{1} \dots \mu_{p+1}} \bigg)~\delta \phi
+ d {\bf \Theta}.
\een
For brevity,
in what follows, we shall denote fields in the underlying theory by $\psi_{\alpha}$,
while their variations by $\delta \psi_{\alpha}$. By virtue of relation (\ref{dl}) we get
the symplectic $(n - 1)$-form
$\Theta_{j_{1} \dots j_{n-1}}[\psi_{\alpha}, \delta \psi_{\alpha}]$, which yields
\be
\Theta_{j_{1} \dots j_{n-1}}[\psi_{\alpha}, \delta \psi_{\alpha}] =
\ep_{\mu j_{1} \dots j_{n-1}} \bigg[
\omega^{\mu} - e^{-{\alpha} \phi} H_{m \nu_{2} \dots \nu_{p+1}}~\delta B_{\nu_{2} \dots \nu_{p+1}}
- \na^{m} \phi~ \delta \phi \bigg],
\ee 
where $\omega_{\mu} = \na^{\alpha} \delta g_{\alpha \mu} - \na_{\mu} 
\delta g_{\beta}{}{}^{\beta}.$
\par 
From Eq.(\ref{dl}) one can see that equation of motion can be read off.
As in Ref.\cite{gao01}, we identify variations of fields with a general coordinate transformations
induced by an arbitrary Killing vector field $\xi_{\alpha}$. In the next step, we calculate
the Noether $(n - 1)$-form with respect to this above mentioned Killing vector, i.e., 
${\cal J}_{j_{1} \dots j_{n-1}} = \ep_{m j_{1} \dots j_{n-1}} {\cal J}^{m}
\big[\psi_{\alpha}, {\cal L}_{\xi} \psi_{\alpha}\big]$. The result of doing that is
\ben
{\cal J}_{j_{1} \dots j_{n-1}} &=& 
d \bigg( Q^{GR} + Q^B \bigg)_{j_{1} \dots j_{n-1}}
+ 2 \ep_{m j_{1} \dots j_{n-1}} \bigg( G^{\delta}{}{}_{\eta} - T^{\delta}{}{}_{\eta}(B, \phi)
\bigg) \xi^{\eta} \\ \nonumber
&+& p~\ep_{m j_{1} \dots j_{n-1}}~ \xi^{d} B_{d \alpha_{2} \dots \alpha_{p+1}}~
\na_{\alpha_{2}} \bigg( e^{-{\alpha} \phi} H^{m \alpha_{2} \dots \alpha_{p+1}} \bigg),
\een
where $Q_{j_{1} \dots j_{n-2}}^{GR}$ yields
\be
Q_{j_{1} \dots j_{n-2}}^{GR} = - \ep_{j_{1} \dots j_{n-2} a b} \na^{a} \xi^{b},
\ee
while $Q_{j_{1} \dots j_{n-2}}^{B}$ has the following form:
\be
Q_{j_{1} \dots j_{n-2}}^{B} = {p \over (p + 1)!} \ep_{m \alpha j_{1} \dots j_{n-1}}
~\xi^{d}~B_{d \alpha_{3} \dots \alpha_{p+1}}~ e^{-{\alpha} \phi} H^{m \alpha \alpha_{3} \dots \alpha_{p+1}}.
\ee
As in Ref.\cite{gao01}, one has in mind that ${\cal J}[\xi] = dQ[\xi] + \xi^{\alpha} {\bf C}_{\alpha}$,
where ${\bf C}_{\alpha}$ is an $(n-1)$-form constructed from dynamical fields, i.e., from
$g_{\mu \nu}$, $(p + 1)$-form field $H^{j_{1} \dots j_{p+1}}$ and dilaton fields. Consequently, one may also identify
$Q_{j_{1} \dots j_{n-1}} = (Q^{GR} + Q^{B})_{j_{1} \dots j_{n-1}}$ with the Noether charge for the considered theory.
Thus, ${\bf C}_{\alpha}$ reduces to
\be
C_{d j_{1} \dots j_{n-1}} = 2 \ep_{m j_{1} \dots j_{n-1}}
\bigg[ G_{d}{}{}^{m} - T_{d}{}{}^{m}(B, \phi) \bigg] +
p~ \ep_{m j_{1} \dots j_{n-1}}~\na_{\alpha_{2}} \bigg( 
e^{-{\alpha} \phi} H^{m \alpha_{2} \dots \alpha_{p+1}} \bigg)
B_{d \alpha_{3} \dots \alpha_{p+1}}.
\ee
The case when ${\bf C}_{\alpha} = 0$ is responsible for the source-free Eqs. of motion. On the other hand,
when this is not the case, it follows directly that we have the following:
\ben
G_{\mu \nu} - T_{\mu \nu}(B, \phi) &=& T_{\mu \nu}(matter) , \\
\na_{\mu_{1}} \bigg( e^{-{\alpha} \phi} H^{\mu_{1} \dots \mu_{p+1}} \bigg)
 &=& j^{\mu_{2} \dots \mu_{p+1}}(matter).
\een
If one further assumes that $(g_{\mu \nu}, B_{\alpha_{1} \dots \alpha_{p}}, \phi)$ are solutions 
of source-free equations of motion and 
$(\delta g_{\mu \nu},~\delta B^{\alpha_{1} \dots \alpha_{p}},~\delta \phi)$
are the linearized perturbations satisfying Eqs. of motion with sources
$\delta T_{\mu \nu}(matter)$ and $ \delta j^{\mu_{1} \dots \mu_{p}}(matter)$, then we reach to the
relation of the form as
\be
\delta  C_{a j_{1} \dots j_{n-1}} = 2 \ep_{m j_{1} \dots j_{n-1}}
\bigg[ \delta T_{a}{}{}^{m}(matter) + p~B_{a \alpha_{3} \dots \alpha_{p+1}}~ 
\delta j ^{m \alpha_{3} \dots \alpha_{p+1}}(matter) \bigg].
\ee
For since the Killing vector field $\xi_{\alpha}$ describes also a symmetry of the background
matter field, one gets the formula for a conserved quantity connected with $\xi_{\alpha}$, namely
\ben \label{hh}
\delta H_{\xi} &=& - 2 \int_{\Sigma}\ep_{m j_{1} \dots j_{n-1}} \bigg[
\delta T_{a}{}{}^{m}(matter) \xi^{a} + 
 p~B_{a \alpha_{3} \dots \alpha_{p+1}}~ 
\delta j ^{m \alpha_{3} \dots \alpha_{p+1}}(matter) \bigg]
 \\ \nonumber
&+& \int_{\p \Sigma}\bigg[
\delta Q(\xi) - \xi \cdot \Theta \bigg].
\een
Let us 
choose $\xi^{\alpha}$ to be an asymptotic time translation $t^{\alpha}$, then
one can conclude that
$M = H_{t}$ and finally obtain 
the variation of the ADM mass
\ben \label{mm}
\alpha~ \delta M &=& - 2 \int_{\Sigma} \ep_{m j_{1} \dots j_{n-1}} \bigg[
\delta T_{a}{}{}^{m}(matter) t^{a} + p~ t^{a} B_{a \alpha_{3} \dots \alpha_{p+1}}~ 
\delta j ^{m \alpha_{3} \dots \alpha_{p+1}}(matter) \bigg] \\ \nonumber
&+& \int_{\p \Sigma}\bigg[
\delta Q(t) - t \cdot \Theta \bigg],
\een
where $\alpha = {n-3 \over n-2}$.
Next, if we take the Killing vector fields $\phi_{(i)}$ which are responsible
for the rotation in the adequate directions, we arrive at the relations for angular 
momenta
\ben \label{jj}
\delta J_{(i)} &=& 2 \int_{\Sigma} \ep_{m j_{1} \dots j_{n-1}} \bigg[
\delta T_{a}{}{}^{m}(matter) \phi_{(i)}^{a} + p~ \phi_{(i)}^{a} B_{a \alpha_{3} \dots \alpha_{p+1}}~ 
\delta j ^{m \alpha_{3} \dots \alpha_{p+1}}(matter) \bigg] \\ \nonumber
&-& \int_{\p \Sigma}\bigg[
\delta Q({\phi}_{(i)} - {\phi}_{(i)} \cdot \Theta \bigg].
\een
Consider now stationary black ring solution to the Eqs. of motion (\ref{m1})-(\ref{m3}).
Let us perturb the black ring by dropping in some matter and assume that in the process
of this action black ring will be not destroyed and settle down to a stationary final state. Just one can
find the changes of the black ring parameters.\\
In order to study the {\it physical process} version of the first law of black ring thermodynamics
let us assume 
that
$(g_{\mu \nu},~B_{\alpha_{1} \dots \alpha_{p}},~\phi)$ are solutions to the source free
Einstein equations with $(p+1)$ form fields and scalar dilaton fields. Suppose,
moreover that the event horizon
the Killing vector field $\xi^{\mu}$ is of the form as
\be
\xi^{\mu} = t^{\mu} + \sum_{i} \Omega_{(i)} \phi^{\mu (i)}.
\ee
In our considerations we shall assume that
$\Sigma_{0}$ is an asymptotically flat
hypersurface which terminates on the event horizon. Then, we take into account 
the initial data on $\Sigma_{0}$ for a linearized perturbations
$(\delta g_{\mu \nu},~ \delta B_{\alpha_{1} \dots \alpha_{p}}, \delta \phi)$
with $\delta T_{\mu \nu}(matter)$ and $\delta j^{\alpha_{2} \dots \alpha_{p+1}}(matter)$. We 
require that $\delta T_{\mu \nu}(matter)$ and $\delta j^{\alpha_{2} \dots \alpha_{p+1}}(matter)$
disappear
at infinity and the initial data for 
$(\delta g_{\mu \nu},~ \delta B_{\alpha_{1} \dots \alpha_{p}}, \delta \phi)$
vanish in the vicinity of the black ring horizon $\cal H$ on 
the hypersurface $\Sigma_{0}$. 
It envisages the fact that for the initial time
$\Sigma_{0}$, the considered black hole is unperturbed. 
Because of the fact that perturbations vanish near the internal boundary $\p \Sigma_{0}$
it follows from relations (\ref{mm}) and (\ref{jj}) that the following is fulfilled:
\ben \label{ppp}
\alpha~ \delta M &-&  \sum_{i} \Omega_{(i)} \delta J^{(i)} = \\ \nonumber
&-& 2 \int_{\Sigma_{0}} \ep_{m j_{1} \dots j_{n-1}} \bigg[
\delta T_{a}{}{}^{m}(matter) \phi_{(i)}^{a} + p~ \phi_{(i)}^{a} B_{a \alpha_{3} \dots \alpha_{p+1}}~ 
\delta j ^{m \alpha_{3} \dots \alpha_{p+1}}(matter) \bigg] \\ \nonumber
&=& \int_{\cH} \gamma ^{\alpha}~k_{\alpha}~\bep_{j_{1} \dots j_{n-1}},
\een
where $\bep_{j_{1} \dots j_{n-1}} = n^{\delta}~\ep_{\delta j_{1} \dots j_{n-1}}$ and
$n^{\delta}$ is the future directed unit normal to the hypersurface $\Sigma_{0}$,
$k_{\alpha}$ is tangent vector to the affinely parametrized null geodesics generators of the event horizon.
Due to the fact of the conservation of current $\gamma^{\alpha}$ and the assumption
that all of the matter falls into the considered black ring we replace in Eq.(\ref{ppp})
$n^{\delta}$ by the vector $k^{\delta}$.
In order to find the integral over the event horizon we take into account the following relation:
\ben \label{cf}
p!~\cL_{\xi} B_{\alpha_{2} \dots \alpha_{p+1}}~\delta j^{\alpha_{2} \dots \alpha_{p+1}}
&-& \xi^{d}~H_{d \alpha_{2} \dots \alpha_{p+1}}~\delta j^{\alpha_{2} \dots \alpha_{p+1}} \\ \nonumber
&=& p~ p! \na_{\alpha_{2}} \bigg(
\xi^{d}~B_{d \alpha_{2} \dots \alpha_{p+1}}
\bigg)~\delta j^{\alpha_{2} \dots \alpha_{p+1}}.
\een
The first term of the left-hand side of Eq.(\ref{cf}) is equal to zero because $\xi_{\alpha}$ is
symmetry of the background solution. Furthermore, let us consider $n$-dimensional Raychauduri 
equation of the form
\be
{d \theta \over d \lambda} = - {\theta^{2} \over (n - 2)} - \sigma_{ij} \sigma^{ij}
- R_{\mu \nu} \xi^{\mu} \xi^{\nu},
\label{ray}
\ee
where $\lambda$ denotes the affine parameter corresponding to vector $k_{\alpha}$, $\theta$ is the expansion and
$\sigma_{ij}$ is shear. They both vanish in the stationary background. 
An inspection of Eq.(\ref{ray}) reveals the fact that
$R_{\alpha \beta} k^{\alpha} k^{\beta} \mid_{\cH} = 0$ which in turn implies the following:
\be
{1 \over 2}k^{\mu} \na_{\mu} \phi~ k^{\nu} \na_{\nu} \phi +
{1 \over 2 p!} e^{- \alpha \phi}
H_{\mu \mu_{2} \dots \mu_{p+1}} H_{\nu}{}{}^{ \mu_{2} \dots \mu_{p+1}} k^{\mu} k^{\nu} \mid_{\cH} = 0.
\ee
Using the fact that $\cL_{k} \phi = 0$,
it is easily seen that,
$H_{\nu}{}{}^{ \mu_{2} \dots \mu_{p+1}} k^{\mu} = 0$. Since 
$H_{\mu \mu_{2} \dots \mu_{p+1}} k^{\mu} k^{\mu_{2}} = 0$, then by asymmetry of $H_{ \mu_{1} \dots \mu_{p+1}}$
it follows that $H_{\mu \mu_{2} \dots \mu_{p+1}} k^{\mu} \sim k_{\mu_{2}} \dots k_{\mu_{p+1}}$.
The pull-back of $H_{\mu}{}{}^{ \mu_{2} \dots \mu_{p+1}} k^{\mu}$ to the event horizon is equal to zero.
Thus,
$\xi^{d}~H_{d \alpha_{2} \dots \alpha_{p+1}}$ is a closed
$p$-form on the horizon. Due to the Hodge theorem (see e.g., \cite{wes81}) it may be rewritten
as a sum of an exact and harmonic form. An exact one does not contribute to the above expression
because of the field Eqs. are fulfilled. The only contribution stems from the harmonic part of
$\xi^{d}~H_{d \alpha_{2} \dots \alpha_{p+1}}$. Having in mind the duality between homology
and cohomology, one can conclude that
there is a harmonic form $\eta$ dual to $n - p- 1$ cycle $\cal S$ in the sense of the equality
of the adequate surface integrals. Just, it follows that the surface term will have the form of
$\Phi_{l}~\delta q_{l}$, where $\Phi_{l}$ is the constant relating to the harmonic part
of $\xi^{d}~H_{d \alpha_{2} \dots \alpha_{p+1}}$ and $\delta q_{l}$ is the variation of a local charge \cite{cop05}.
These allow one to write down the following:
\be
\alpha~ \delta M -  \sum_{i} \Omega_{(i)} \delta J^{(i)} 
+ \Phi_{l}~\delta q_{l} =
2 \int_{\cH}
\delta T_{\mu}{}{}^{\nu} \xi^{\mu} k_{\nu}.
\label{rh}
\ee
In order to calculate the right-hand side of Eq.(\ref{rh}),
one can use the same procedure as described in Refs.\cite{gao01,rog02,rog05}. Namely,
considering $n$-dimensional Raychauduri Eq. and 
using the fact that
the null generators of the event horizon of the perturbed black ring coincide with
the null generators of the unperturbed stationary black ring, lead to the conclusion that
\be
\kappa~ \delta A = \int_{\cal H} \delta
T^{\mu}{}{}_{\nu}(matter) \xi^{\nu} k_{\mu},
\ee
where $\kappa$ is the surface gravity.\\
In the light of what has been shown we obtained the
{\it physical process} version of the first law of black ring
mechanics in Einstein gravity with
additional
$(p+1)$-form field strength and dilaton fields. It is of the same form as known from Ref.\cite{cop05}, namely
\be
\alpha~ \delta M - \sum_{i} \Omega_{(i)} \delta J^{(i)} 
+ \Phi_{l}~\delta q_{l}
 = \kappa ~\delta {\cal A}.
\ee
We finally remark that a proof of {\it physical process} version of the first law of
thermodynamics for $n$-dimensional black rings also provides support for cosmic censorship.

\noindent
{\bf Acknowledgements:}\\
MR was supported in part by the Polish Ministry of Science and Information Society Technologies grant 1 P03B 049 29.


\end{document}